\documentclass[prb,aps,onecolumn,showpacs,floats]{revtex4}
\usepackage{epsfig,bm,epsf,graphics}
\begin{document}
\draft
\title{Potts-Percolation-Gauss Model of a Solid}
\author{Miron Kaufman$^{a}$\footnote{ Corresponding author, E-mail:m.kaufman@csuohio.edu } and H. T. Diep$^b$}
\address{
$^a$ Department of Physics, Cleveland State University, Cleveland, OH.44115, USA\\
$^b$ Laboratoire de Physique Th\'eorique et Mod\'elisation,
CNRS-Universit\'e de Cergy-Pontoise, UMR 8089\\
2, Avenue Adolphe Chauvin, 95302 Cergy-Pontoise Cedex, France\\}

\begin{abstract}
We study a statistical mechanics model of a solid.  Neighboring
atoms are connected by Hookian "springs".  If the energy is larger
than a threshold the "spring" is more likely to fail, while if the
energy is lower than the threshold the spring is more likely to be
alive. The phase diagram and thermodynamic quantities, such as
free energy, numbers of bonds and clusters, and their
fluctuations, are determined using renormalization-group and
Monte-Carlo techniques.
\end{abstract}

\maketitle
\section{Introduction}
The mechanical properties of solids, such as the mechanical
failure, are topics of considerable
interest.\cite{Arcangelis,Beale,Bolander,Buxton,Yanay}  In this
paper we analyze an equilibrium statistical mechanics
model\cite{Blumberg} of a solid.  In previous
calculations\cite{Kaufman96}we went beyond the ideal Hooke law for
springs by using the realistic anharmonic energy versus atomic
distance developed and tested extensively by Ferrante and his
collaborators.\cite{Rose}  We found that the phase diagram
exhibits universal features when the temperature and the stress
are appropriately scaled.  Those calculations were mean-field in
character as we assumed that all springs have the same strain.  In
this paper we evaluate the role of thermal fluctuations by using
renormalization-group and Monte-Carlo simulations.
    The model is defined in Section II.  We view the solid as a
collection of harmonic springs.  If the energy of such a spring is
larger than a threshold, the spring is likely to
fail.\cite{Hassold} Assuming that the relaxation times are short
compared to the measurement time, we use equilibrium statistical
mechanics to compute the various thermodynamic quantities. When
the elastic energy is not too large, the partition function for
the harmonic "springs" defined on percolation clusters can be
mapped into a Potts model. \cite{Potts,Wu} This model is quite
similar to the annealed Ising model on percolation
cluster.\cite{Kaufman94,Scholten}
    In Section III we present numerical results based on the renormalization-group Migdal-Kadanoff scheme
which, as first shown by Berker\cite{Berker}, provides exact
solutions of statistical models on hierarchical lattices
\cite{Kaufman81,Kaufman84,Erbas} and on small world
nets.\cite{Hinczewski,Rozenfeld}  Monte-Carlo simulations are
presented in Section IV.  Our concluding remarks are found in
Section V.

\section{Model }

The energy of a "spring" $(i,j)$ is given by the Hooke law:

\begin{equation}
H_{ij}=-E_C+\frac{k}{2}(r_i-r_j)^2
\end{equation}
where $r_i$ is the displacement vector from the equilibrium
position of atom $i$, measured in units of the lattice constant
$a$, $E_C$ is the cohesive energy, and $k$ is elastic constant. If
the energy of the spring is larger than the threshold energy $E_0$
the "spring" is more likely to fail than to be alive. $p$ is the
probability that the "spring" is alive and $1- p$ the probability
that the "spring" breaks.  We assume its dependence on energy to
be given by the Boltzmann weight

\begin{equation}
\frac{p}{1-p}=e^{-\frac{{\cal
H}-E_0}{k_BT}}=we^{-\frac{K}{2}(r_i-r_j)^2}
\end{equation}
where: $K= k/k_BT$ and $w=e^{\frac{E_C+E_0}{k_BT}}$.
    We allow for correlations between failing
events by using the Potts\cite{Potts} number of states $q$, which
plays the role of a fugacity controlling the number of
clusters\cite{Kaufman84b,Hu}: if $q >> 1$ there is a tendency of
forming many small clusters while if $q << 1$ there is a tendency
to form a few large clusters. If $q = 1$ springs fail
independently of one another, i.e. random percolation
problem\cite{Fortuin}. The partition function is obtained by
summing over all possible configurations of bonds arranged on the
lattice

\begin{equation}
Z=\sum_{config}q^cw^B Z_{\mbox{elastic}}^{\mbox{config}}
\end{equation}

$C$ is the number of clusters, including single site clusters, and
$B$ is number of live "springs". The restricted partition function
associated with the elastic energy for a given configuration of
bonds (live "springs") is

\begin{equation}
Z_{\mbox{elastic}}^{\mbox{config}}=Tr_r e^{
-\frac{H_{\mbox{elastic}}}{k_BT}}
\end{equation}

\begin{equation}\label{H1}
-\frac{H_{\mbox{elastic}}}{k_BT}=\sum_{<i,j>}\frac{K}{2}(r_i-r_j)^2
\end{equation}

In Eq. (\ref{H1}) the sum is over all live "springs".

    By using the Kasteleyn-Fortuin expansion\cite{Fortuin} for Potts model we can rewrite the partition function as

\begin{equation}
Z=Tr_\sigma Tr_r e^{-\frac{{\cal H}}{k_BT}}
\end{equation}

The Hamiltonian is

\begin{equation}\label{H}
-\frac{{\cal H}}{k_BT}=\sum_{<i,j>} [J_1 \delta(\sigma_i,\sigma_j)
-\frac {J_2}{2}\delta(\sigma_i,\sigma_j) (r_i-r_j)^2]
\end{equation}
where $\sigma_i$ is a Potts spin taking $q$ values.  This mapping
is a Gaussian approximation valid when, on the right hand side of
Eq. \ref{H}, the elastic energy is small compared to the first
energy contribution.  The coupling constants $J_1$ and $J_2$ are
related to the original parameters, $w$ and $K$, as follows:

\begin{eqnarray}
J_1&=&\ln (1+w)\label{J1}\\
J_2&=&K\frac{w}{w+1}\label{J2}
\end{eqnarray}

    The free energy per bond is: $f = \ln Z/N_{\mbox{bonds}}$.  The derivatives of the free energy $f$
 with respect to the parameters $w$, $K$, and $q$ provide respectively the number of live "springs" $b$,
 the elastic energy $E_{\mbox{elastic}}$ and the number of clusters $c$, each normalized by the total number of lattice bonds:

\begin{eqnarray}
E_{\mbox{elastic}}&=&-K\frac{\partial f}{\partial K}\\
b&=&w\frac{\partial f}{\partial w}\\
c&=&q\frac{\partial f}{\partial q}
\end{eqnarray}

The derivatives of those densities, $E_{\mbox{elastic}}$, $b$, and
$c$, with respect to the model parameters provide in turn the
fluctuations (variances) of those quantities:

\begin{eqnarray}
\Delta E^2_{\mbox{elastic}}&=&-K\frac{\partial E_{\mbox{elastic}}}{\partial K}+E_{\mbox{elastic}}\\
\Delta b^2&=&w\frac{\partial b}{\partial w}\\
\Delta c^2&=&q\frac{\partial c}{\partial q}
\end{eqnarray}

\section{Renormalization Group}

 The Migdal-Kadanoff recursion
equations\cite{Migdal,Kadanoff} for $d$ dimensions are:
$Z_{i,j}=(Tr_kZ_{i,k}Z_{k,j})^L$, where $L=2^(d-1)$.  We assume
each atom coordinate varies in the interval (-1/2, 1/2), where the
equilibrium lattice constant is 1. After also using the Gaussian
approximation (small elastic energy) we get:

\begin{eqnarray}
w'&=&[1+U(w,K,q)]^L-1\label{WK}\\
K'w'&=&K\frac{L}{2}[1+U(w,K,q)]^{L-1}U(w,K,q)\label{WK1}
\end{eqnarray}
where $L=2^{d-1}$ and

\begin{equation}
U(w,K,q)=\frac{w^2
erf(\sqrt{K/4})}{q\sqrt{K/\pi}+\sqrt{8}w.erf(\sqrt{K/8})}
\end{equation}

The recursion equations (\ref{WK})-(\ref{WK1}) represent the
Gaussian approximation of the exact solutions for hierarchical
lattices.  Since this scheme is realizable, the convexity of the
free energy is preserved\cite{Kaufman83}, and thus reasonable
expectations, such as positivity of energy fluctuations, are
fulfilled.
    The renormalization group flows are governed by the
following fixed points at $K = 0$ (pure Potts model): i. $w = 0$
(non-percolating live "springs"), ii. $w = 8$ (percolating network
of live "springs"), iii. $w = w_c$ (Potts critical point).  A
stability analysis at the Potts critical point, ($K = 0$, $w =
w_c$) yields the two eigenvalues: i. the thermal eigenvalue
$\Lambda_1$ (for the direction along the $K= 0$ axis) is always
larger than 1, meaning the $w- w_c$ is a relevant field; ii. The
other eigenvalue $\Lambda_2$ is associated with the flow along the
$w = w_c$ line away from the pure model ($K = 0$).  For $d = 2$,
$\Lambda_2 < 1$ for all $q$.  This means that there is a line of
points in the $(w, K)$ flowing into, and thus is in the same
universality class as, the pure Potts critical point ($w_c, 0)$.
For $d = 3$ on the other hand, $\Lambda_2 < 1$ for $q < 109$, but
$\Lambda_2 > 1$ for $q > 109$.  There exists another fixed point
at $(w^*, K^*)$ which has both eigenvalues larger than unity for
$q < 109$, and becomes stable in one direction for $q > 109$.
Thus in $d = 3$, for large enough $q$, the elastic constant $K$
becomes a relevant field changing the universality class of the
model from the pure Potts criticality to a new one, Potts-elastic.
However, in view of the Gaussian approximation used to derive the
recursion equations, we view this as only an indication of a
possible new universalty class that warrants further study.


\begin{figure}[t!]
\centering
\includegraphics[width=5.0in]{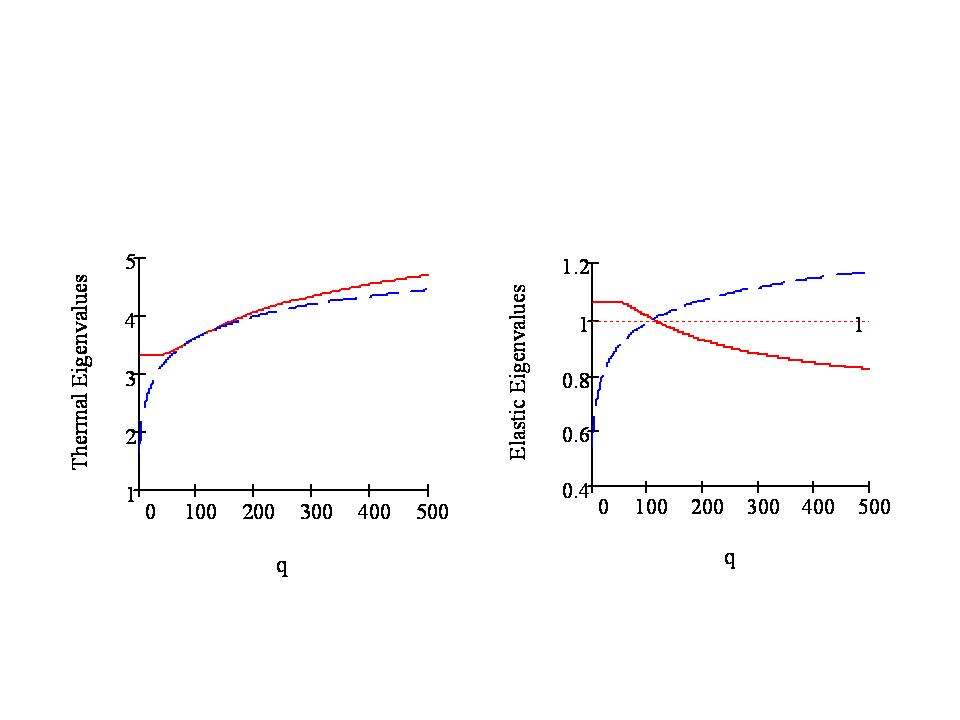}
\caption{Thermal and Elastic Eigenvalues in $d=3$.} \label{fig:1}
\end{figure}

The phase diagram for any given $q$, in the $(w, K)$ plane shows
two phases: I. solid with a percolating network of live "springs",
II. crumbling solid with mostly "failed" springs.  The two phases
are separated by a critical line in the universality class of the
$q$-state Potts model (for $d = 2$ for all $q$, for $d = 3$ for $q
< 109$).


\begin{figure}[t!]
\centering
\includegraphics[width=5.0in]{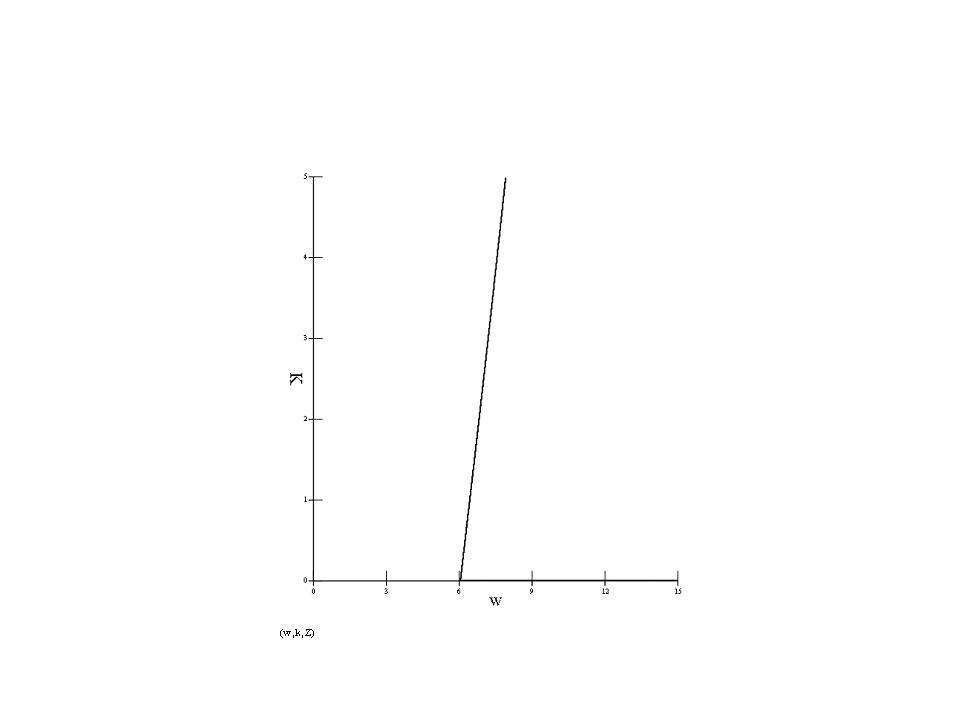}
\caption{Phase diagram for $q = 10$ and $d = 2$, in the plane ($w,
K$).}\label{fig:2}
\end{figure}

Note that when increasing the elastic constant $K$, one needs a
higher $w$ to establish the solid phase. This is due to the fact
that increasing the elastic energy increases the probability for
the "spring" to fail.

    The free energy $f = \ln Z/N_B$ , where $N_B$ is the number of lattice edges, is:

\begin{equation}
f=\sum_{n=1}^{\infty}\frac{C_n}{(2L)^n}
\end{equation}
where
\begin{equation}
C=\ln[q+w\sqrt{\frac{8\pi}{K}} erf (\sqrt{\frac{K}{8}})]^L
\end{equation}
and $L = 2^{d -1}$.  Using the free energy we can compute the
number of live "springs" $b$, the number of clusters $c$, the
elastic energy $E_{\mbox{elastic}}$, and their fluctuations
(variances). Each of those quantities is scaled by the total
number of lattice edges $N_B$.
    In Figure 3 we show the elastic energy variation with K and w for two different q values.
As expected the elastic energy increases monotonically with K and
w starting at zero at K = 0 and at w = 0.


\begin{figure}[t!]
\centering
\includegraphics[width=5.0in]{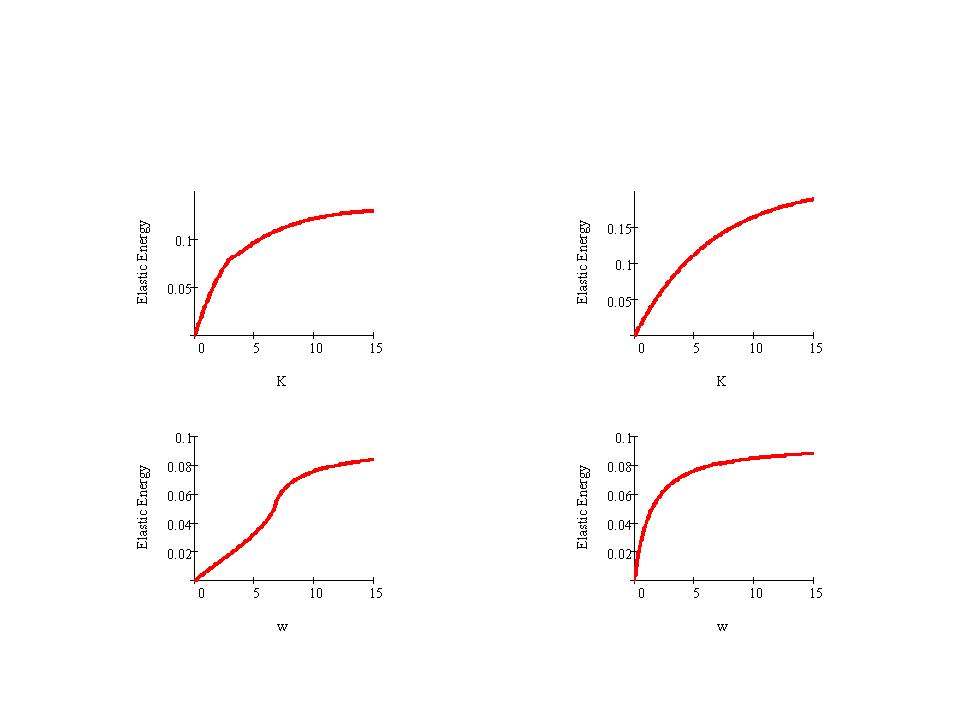}
\caption{ Elastic energy vs $K$ and $w$. Left column $q = 10$;
Right column $q = 1$.} \label{fig:3}
\end{figure}

    The number of live "springs" increases with w and decreases with K, as shown in Figure 4.


\begin{figure}[t!]
\centering
\includegraphics[width=5.0in]{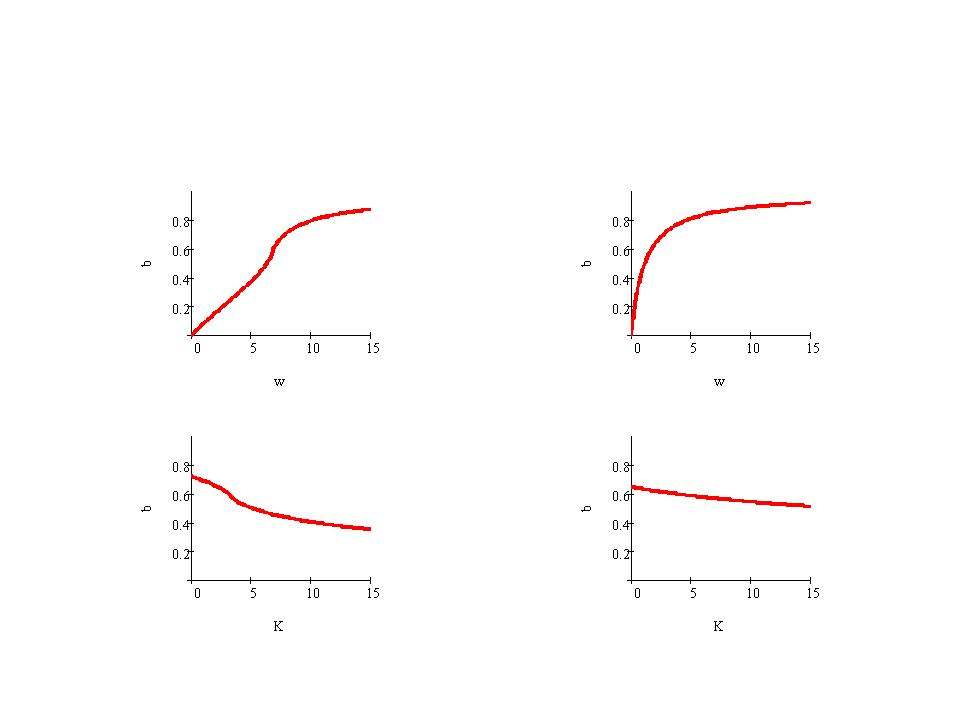}
\caption{ Number of live springs vs. $w$ and $K$, respectively.
Left column $q = 10$; Right column $q = 1$.} \label{fig:4}
\end{figure}

    We also estimate the squared mean elongation (in units of lattice constant $a$) of live "sprins"
by using the number of live springs, $b$, and the elastic energy:

\begin{equation}
\Delta r^2=\frac{2}{K}\frac{E_{\mbox{elastic}}}{b}
\end{equation}

One can use the classical Lindemann model of
melting\cite{Lindemann}, $\Delta r^2 = 0.01$, to estimate the
melting temperature of our model solid.


\begin{figure}[t!]
\centering
\includegraphics[width=5.0in]{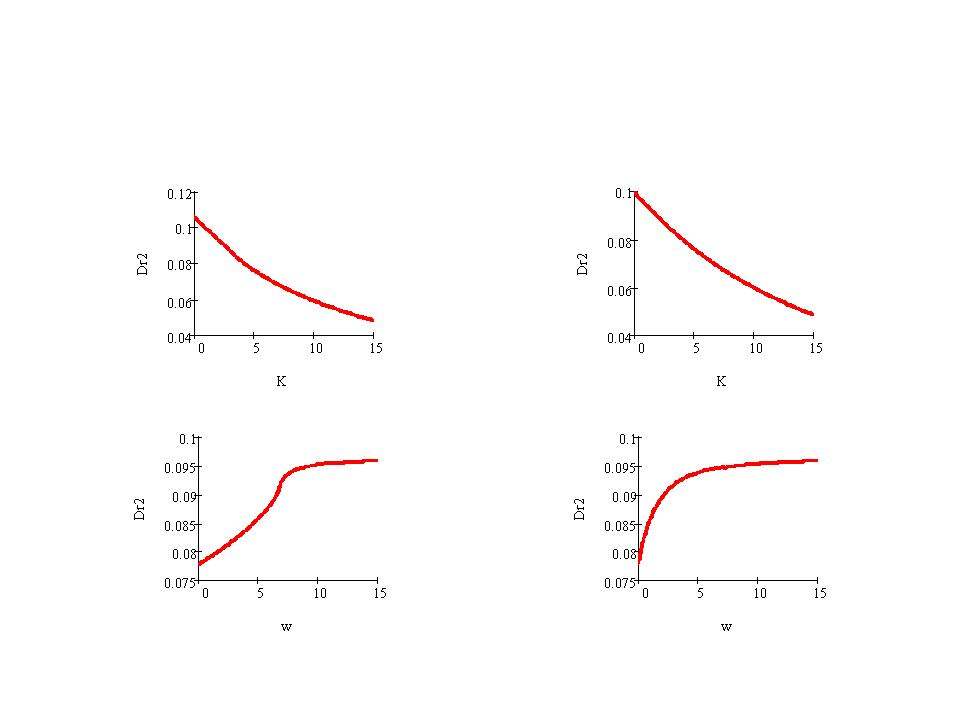}
\caption{ Square of live "spring" elongation vs. $w$ and $K$,
respectively. Left column $q = 10$; Right column $q = 1$.}
\label{fig:5}
\end{figure}

    In the limit $w = 0$, the number of clusters is equal to the number of sites. Hence $c$ approaches the inverse of the coordination number,
    which for the diamond hierarchical lattice, correponding to the Migdal-Kadanoff scheme for $d =2$, is\cite{Kaufman84}:
$c = 2/3$, consistent with Figure 6.


\begin{figure}[t!]
\centering
\includegraphics[width=5.0in]{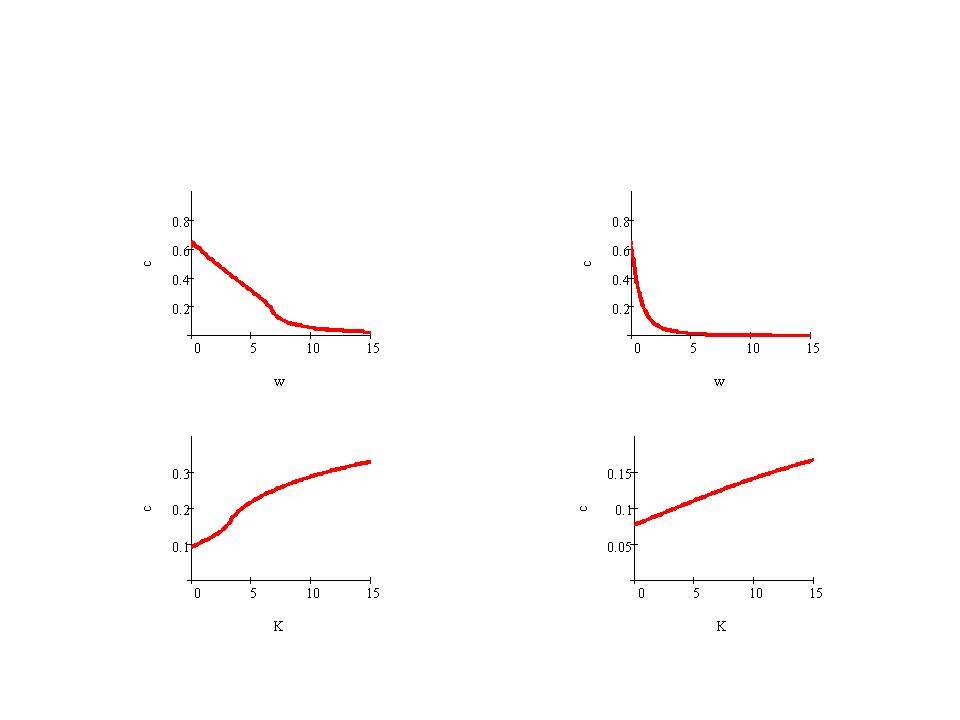}
\caption{ Number of clusters vs. $w$ and $K$, respectively.  Left
column $q$ = 10; Right column $q$ = 1.} \label{fig:6}
\end{figure}

    In Figure 7 we show the elastic energy fluctuations and the number of
live "springs" fluctuations as functions of $K$ and $w$, for $q =
10$ and $q$= 1 respectively.  Since the exponent $\alpha$ is
positive for $q = 10$ and negative for $q = 1$, a divergence is
apparent in the $q = 10$ critical point $K = 2$, $w = 7.3$.


\begin{figure}[t!]
\centering
\includegraphics[width=5.0in]{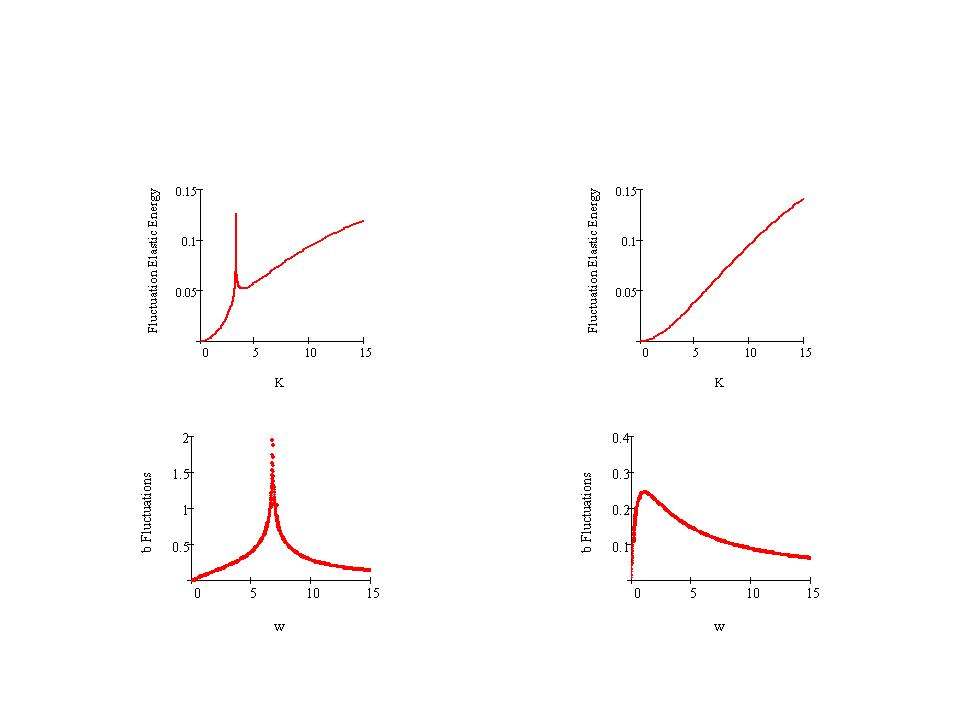}
\caption{ Fluctuations (variances) in elastic energy and number of
live "springs" vs. $K$ and $w$, respectively.  Left column $q$ = 10;
Right column $q$ = 1.} \label{fig:7}
\end{figure}

    The number of clusters $c$ increases monotonically with the conjugated fugacity $q$, starting at $c$ = 0 at $q$ = 0, as shown in Figure 8.
The fluctuations in c exhibit a divergence at the critical point
$q$ = 10,  $K$= 2, $w$ = 7.3 where the critical exponent $\alpha$
is positive.


\begin{figure}[t!]
\centering
\includegraphics[width=5.0in]{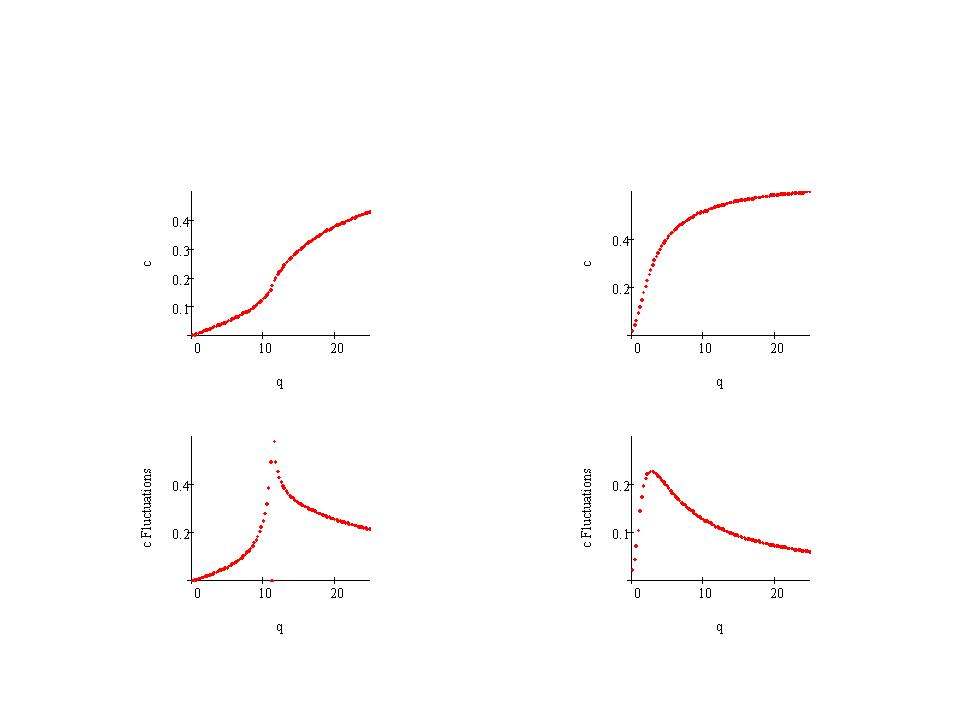}
\caption{ Number of clusters and their fluctuations (variances) vs.
$q$.  Left column $w = 7.3$, $K = 2$; Right column $w = 1.857$, $K =
2$.} \label{fig:8}
\end{figure}

    In Figure 9 we show fluctuations in elastic energy, in number of
bonds, and number of clusters against $w$, $K$, and $q$
respectively, for $d = 3$ and $q = 20$. One can notice the lack of
symmetry in the divergence which is a characteristic of 3$d$
criticality\cite{Kaufman84a,Chase}. By contrast in $2d$ the
divergences are symmetric (see Figure 7).  This is related to the
duality transformation \cite{Kaufman84a,Kaufman84b}. The critical
point exponent at $q = 20$, $w = 3.9$, $K = 2$ is $\alpha =
0.015$.


\begin{figure}[t!]
\centering
\includegraphics[width=5.0in]{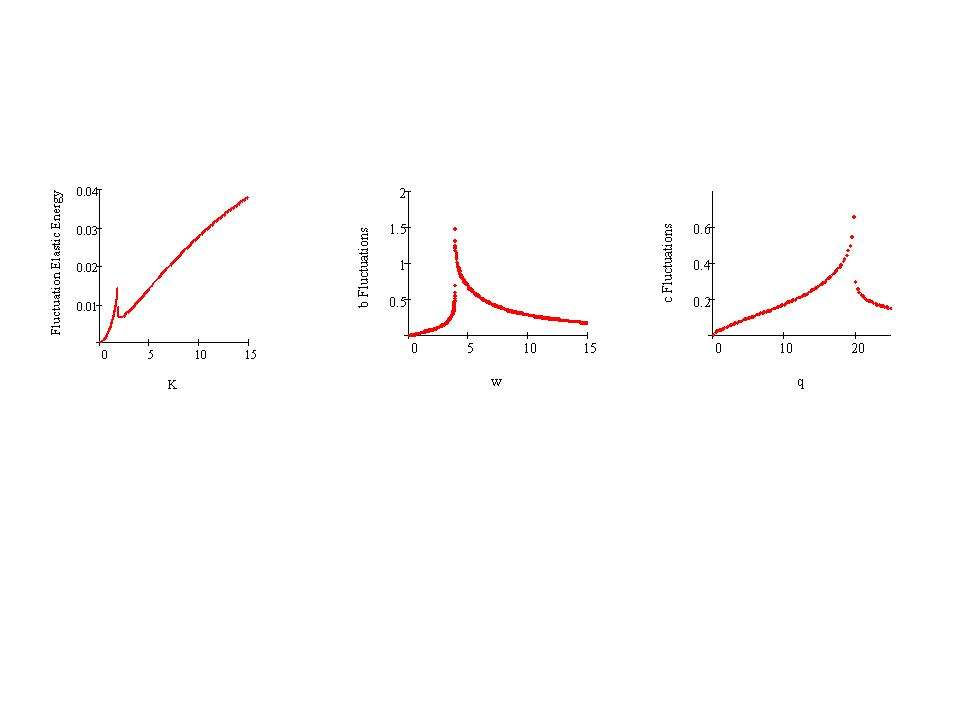}
\caption{ Fluctuations of elastic energy, number of live "springs",
and number of clusters vs. $k$, $w$, and $q$, respectively, for $d$
= 3.} \label{fig:9}
\end{figure}

\section{Monte Carlo Simulations}

In this section, we use  for Monte Carlo simulation the following
Hamiltonian taken from Eq. (7)
\begin{equation}\label{HMC}
{\cal H}=-I_1\sum_{<i,j>} \delta(\sigma_i,\sigma_j) +\frac
{I_2}{2}\sum_{<i,j>}\delta(\sigma_i,\sigma_j) (r_i-r_j)^2
\end{equation}
where $I_1$ and $I_2$ are renormalized parameters.  Of course, one
has $J_1=I_1/k_BT$ and $J_2=I_2/k_BT$.

  In the case of two dimensions $d=2$, we consider a square lattice of
size $N\times N$ where $N=40,60,80,100$.  Each lattice site is
occupied by a $q-$state Potts spin.  We use periodic boundary
conditions. Our purpose here is to locate the phase transition
point and establish the phase diagram.  To this end, a simple
heat-bath Metropolis algorithm is sufficient.\cite{Binder}  The
determination of the order of the phase transition and the
calculation of the critical exponents in the second-order phase
transition region need more sophisticated Monte Carlo methods such
as histogram techniques.\cite{Ferren}  These  are left for a
future study.

The simulation is carried out as follows.   For each set of
$(I_1,I_2)$, we equilibrate the system at a given temperature $T$
during $10^6$ Monte Carlo sweeps (MCS) per spin before averaging
physical quantities over the next $10^6$ MCS.  In each sweep, both
the  spin value and the spin position are updated according to the
Metropolis criterion. The calculated physical quantities are the
internal energy per spin $E$, the specific heat $C_v$ per spin,
the Potts order parameter $Q$ and the susceptibility per spin
$\chi$. For a $q$-state Potts model, $Q$ is defined as

\begin{equation}
Q=\frac{q \max (Q_1,Q_2,..., Q_q)-1}{q-1} \nonumber
\end{equation}
where $Q_i=\frac{n_i}{N^2}$ ($i=1,...,q)$, $n_i$ being the number
of sites having $q_i$.

\begin{figure}[t!]
\centering
\includegraphics[width=2.0in,angle=-90]{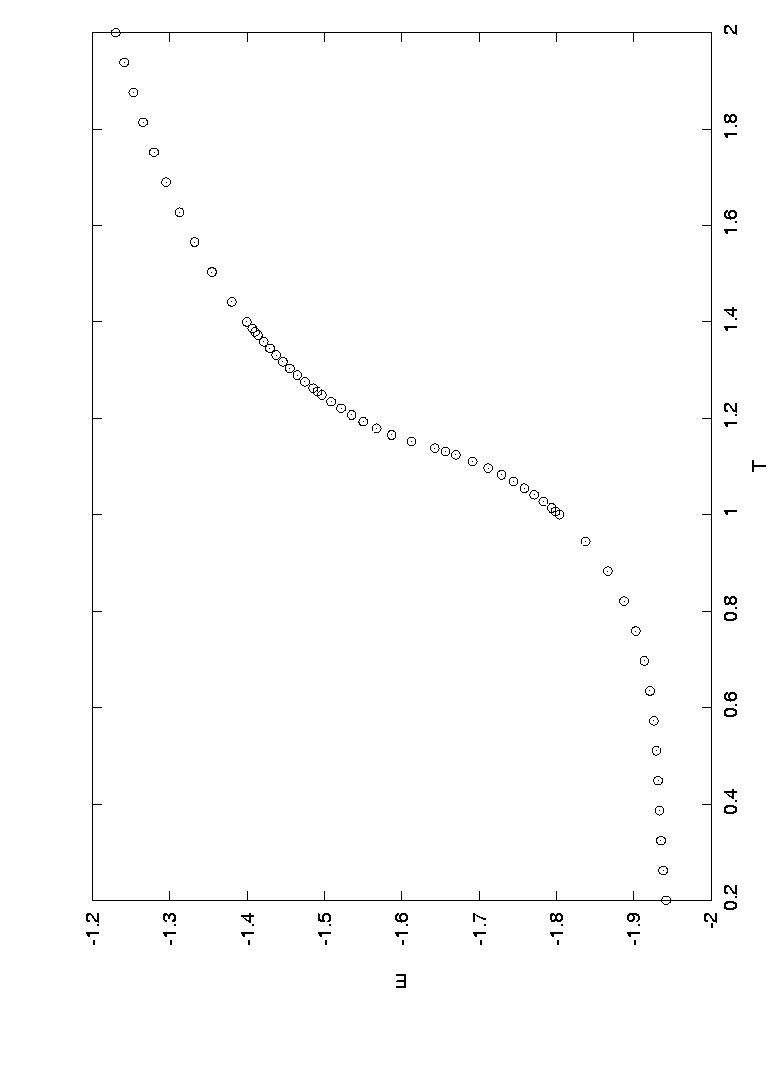}
\includegraphics[width=2.0in,angle=-90]{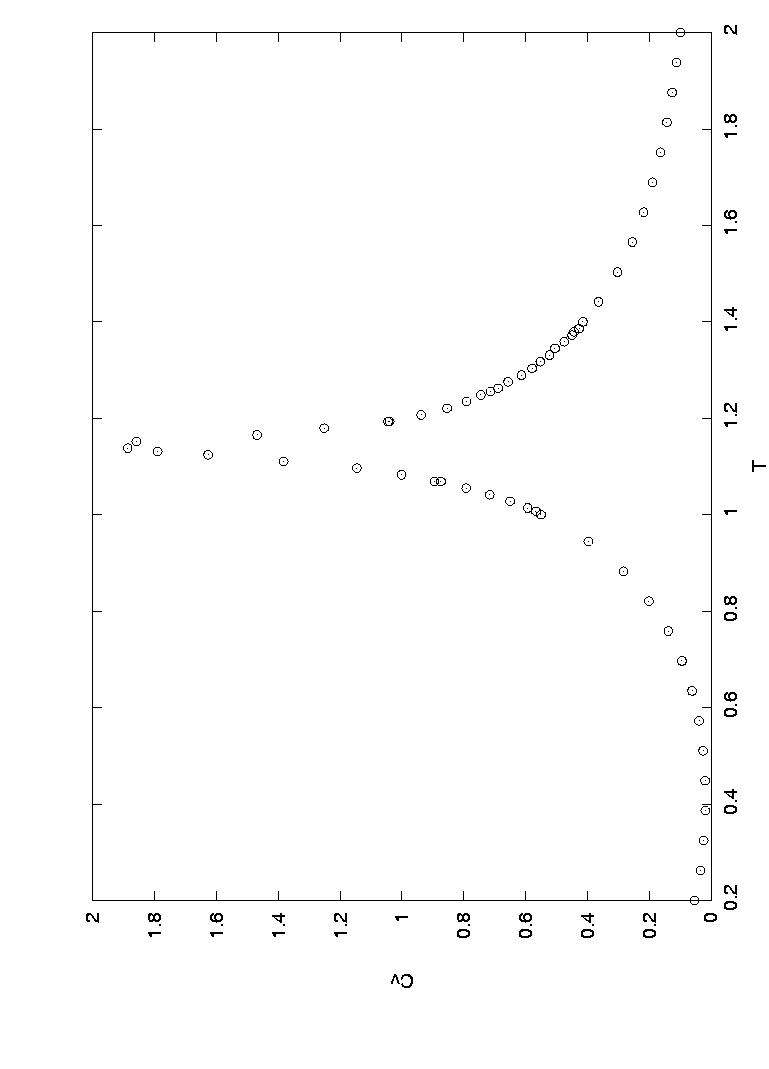}
\caption{ Energy per spin $E$ (left) and specific heat per spin
$C_v$ (right) vs temperature $T$ for $q=2$, $I_2 = 0.5$ with $N=60$
and $I_1 = 1$. } \label{fig:E}
\end{figure}

\begin{figure}[t!]
\centering
\includegraphics[width=2.0in,angle=-90]{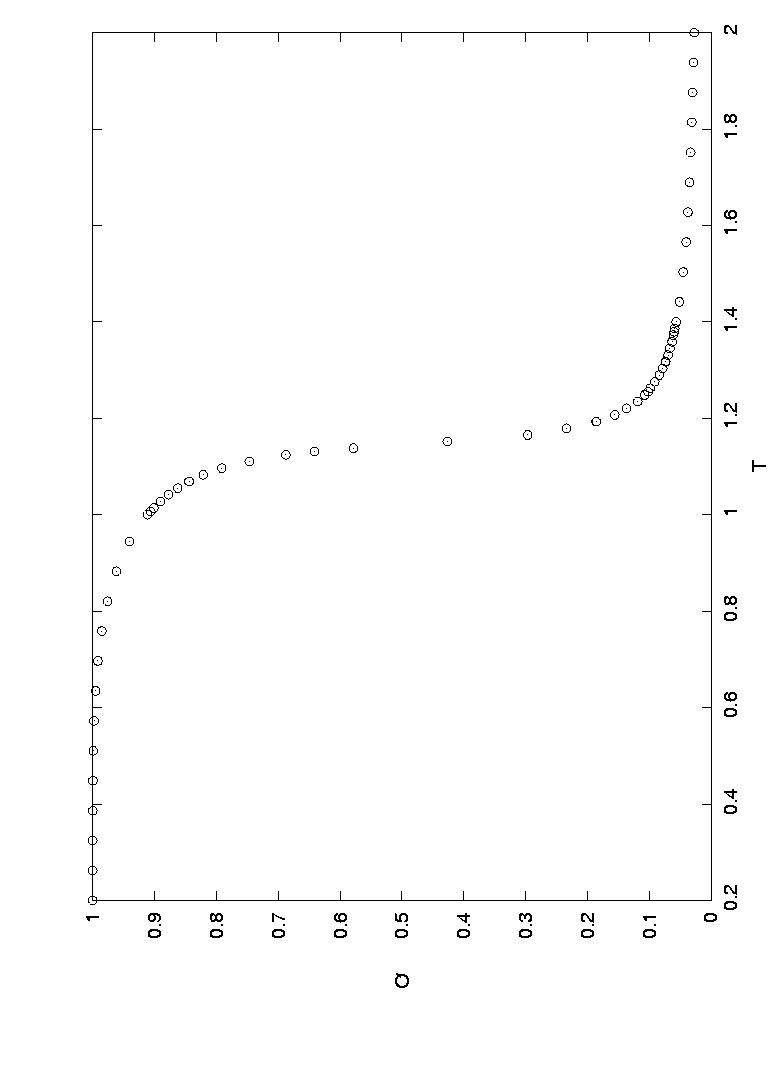}
\includegraphics[width=2.0in,angle=-90]{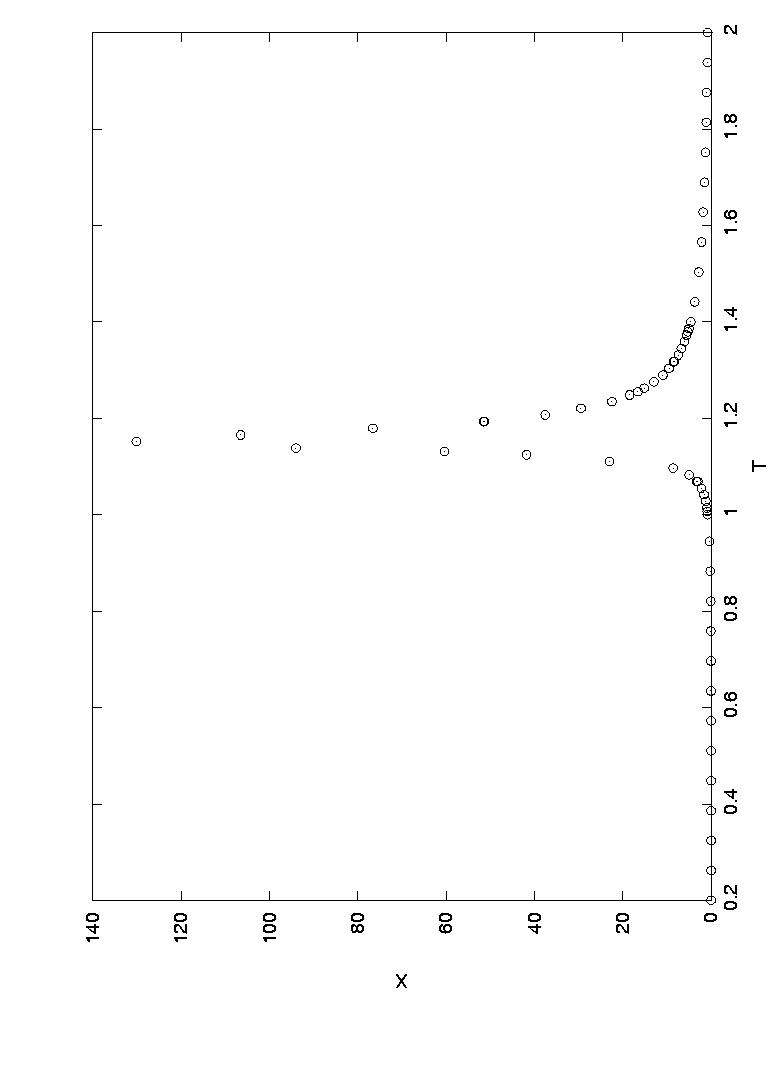}
\caption{ Order parameter $Q$ (left) and susceptibility $\chi$
(right) vs temperature $T$ for $q=2$, $I_2 = 0.5$ with $N=60$ and
$I_1 = 1$. } \label{fig:Q}
\end{figure}

Let us show first in Fig. \ref{fig:E} and  Fig. \ref{fig:Q} the
energy, the specific heat, the order parameter and the
susceptibility in the case where $q=2$, $I_1=1$ and $I_2=0.5$.


These figures show a phase transition at $T_c=1.130\pm 0.005$.
Note that the size effects for $N=40$, 60, 80 and 100 are not
significant and are included in the error estimation. Simulations
have been carried out also for the following sets $(I_1=1,
I_2=0.2)$, $(I_1=1, I_2=0.8)$ and $(I_1=1, I_2=1)$.  The results
show that the transition temperature $T_c$ does not change
significantly with this range of $I_2$.  $T_c$ depends only on the
main $I_1$ term.

To compare with the results from the renormalization group
calculation of the previous section, we have to use
$J_1=I_1/k_BT$, $J_2=I_2/k_BT$ and Eqs. (\ref{J1}-\ref{J2}) to
convert $J_1$ and $J_2$ into $K$ and $w$.  One has

\begin{eqnarray}
w&=&\exp(\beta I_1) - 1 \label{w1}\\
K&=&\frac {I_2}{k_BT}\frac{1}{1-\exp(-\beta I_1)}\label{K1}
\end{eqnarray}
where $\beta=1/k_BT$.

\begin{figure}[t!]
\centering
\includegraphics[width=2.0in,angle=-90]{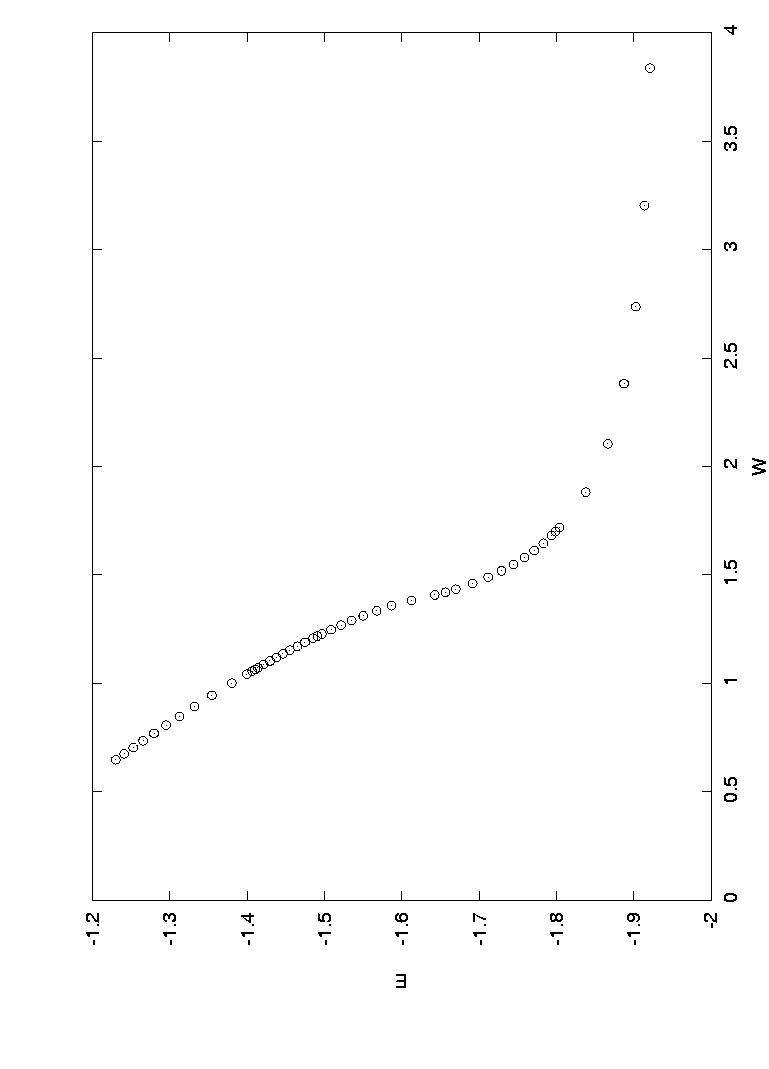}
\includegraphics[width=2.0in,angle=-90]{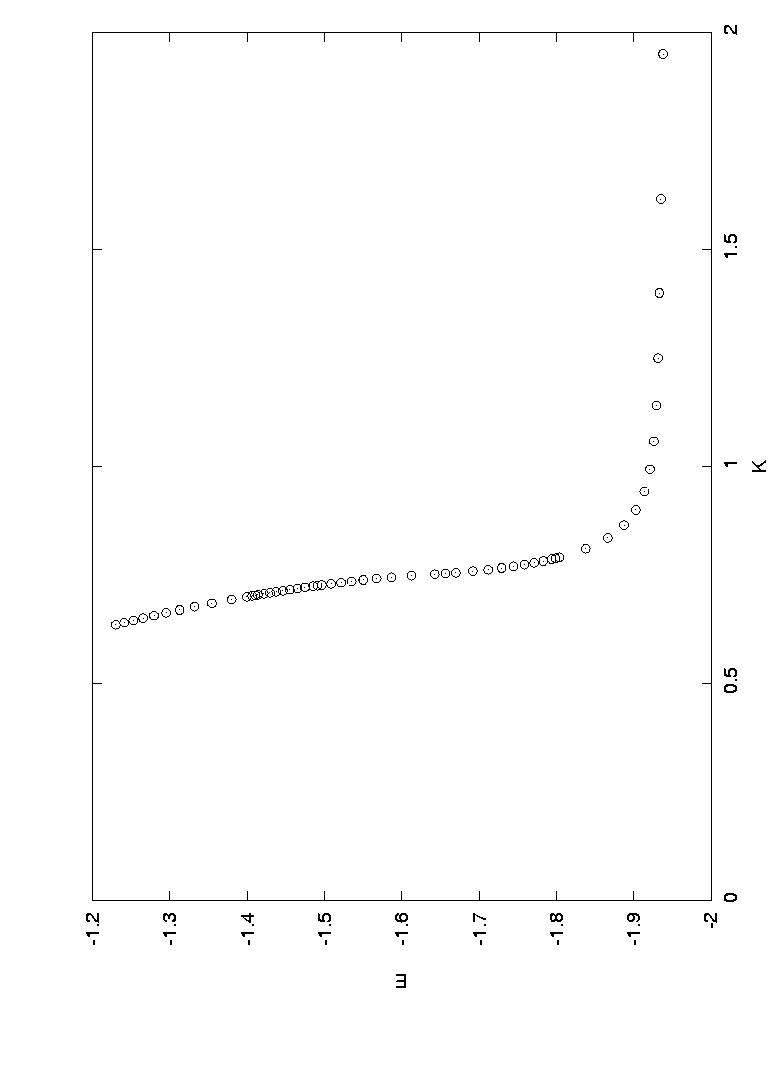}
\caption{ Internal energy $E$ versus $w$ (left) and versus $K$
 (right) for
$q=2$, $I_2 = 0.5$ with $N=60$ and $I_1 = 1$. } \label{fig:EWK}
\end{figure}

Figure \ref{fig:EWK} shows the internal energy $E$ as functions of
$w$ and $K$ for $I_1=1$ and $I_2=0.5$.  The transition is found at
$w^*=1.421\pm 0.002$ and $K^*=0.753\pm0.002$.    We observe that
only $K^*$ varies with $I_2$, not $w^*$ as expected from Eqs.
(\ref{w1})-(\ref{K1}).  We have $K^*=0.301, 1.205$ and $1.506$ for
$I_2=0.2, 0.8$ and 1, respectively.


\section{ Conclusions}

We have studied a model of a solid made of "springs" that are live
and harmonic or failed. The "springs" can fail with a probability
that increases with the energy.  Our renormalization-group
analysis suggests that elastic perturbations on the
Potts-percolation model are irrelevant for all $q$ in two
dimensions, and for small enough $q$ in three dimensions. The
renormalization-group predictions must be viewed as only
indicative, in view of the following known limitations.  The
simple Migdal-Kadanoff renormalization group fails to predict
correctly the first-order transitions for $d = 2$, $q > 4$, and
for $d = 3$, $q > 2$. Furthermore our recursion equations are
valid for small elastic energy.  Monte Carlo simulations are
needed to further our understanding of the model. Monte Carlo
results in $d=2$ show that the phase transition does not depend on
$I_2$ both on the value of the transition temperature and on the
transition order. This is physically in agreement with the fact
that in two dimensions the melting does not take place at finite
temperature. At least in $d=2$, one can say that the phase
transition is solely due to the first term of the Hamiltonian
(\ref{HMC}).  We note in passing that in the Ising case, the
question of whether the elastic interaction affects or not the
Ising universality class has been
investigated.\cite{Bergmann,Fisher68} No definite conclusions have
been reached.\cite{Diep,Landau} It would be therefore interesting
in the future to perform Monte Carlo simulations for $d=3$ and for
large $q$ to investigate the effect of elastic interaction on the
criticality and on cross-over from second to first-order.

{}


\begin{thebibliography}{}

\bibitem{Arcangelis}
 L. De Arcangelis, A. Hansen, H. J. Hermann, S. Roux,
Phys. Rev. B {\bf 40}, 877 (1989).

\bibitem{Beale} P. D. Beale, D. J. Srolovitz, Phys. Rev. B {\bf 37}, 5500 (1988).

\bibitem{Bolander} J. E. Bolander, N. Sukumar, Phys. Rev. B {\bf 71}, 094106 (2005).

\bibitem{Buxton} G. A. Buxton, R. Verberg, D. Jasnow, A. C. Balazs, Phys. Rev. E {\bf 71}, 056707 (2005).

\bibitem{Yanay} Y. Yanay, A. Goldsmith, M. Siman, R. Englman, Z. Jaeger, J. Appl.Phys. {\bf 101}, 104911
(2007).

\bibitem{Blumberg} R. L. Blumberg Selinger, Z. G. Wang, W. M. Gelbart, A.
Ben-Shaul, Phys. Rev. A {\bf 43}, 4396 (1991).

\bibitem{Kaufman96} M. Kaufman, J. Ferrante, NASA Tech. Memo. 107112 (1996).

\bibitem{Rose} J. H. Rose, J. Ferrante, J. R. Smith, Phys. Rev. Lett. {\bf 47} , 675
(1981); J. H. Rose, J. R. Smith, F. Guinea, J. Ferrante, Phys.
Rev. B {\bf 29}, 2963 (1984); J. Ferrante, J. R. Smith, Phys. Rev.
B {\bf 31}, 3427 (1985).

\bibitem{Hassold} G. N. Hassold and D. J. Srolovitz, Phys. Rev. B {\bf 39}, 9273
(1989).

\bibitem{Potts} R. B. Potts, Proc. Camb. Phil. Soc. {\bf 48}, 106 (1952).

\bibitem{Wu} F. Y. Wu, Rev. Mod. Phys. {\bf 54}, 235-268 (1982).

\bibitem{Kaufman94} M. Kaufman and J. E. Touma, Phys. Rev. B {\bf 49}, 9583 (1994).

\bibitem{Scholten} P. D. Scholten and M. Kaufman, Phys. Rev. B {\bf 56}, 59 (1997).

\bibitem{Berker} A. N. Berker and S. Ostlund, J. Phys. C {\bf 12}, 4961-4975 (1979).

\bibitem{Kaufman81} M. Kaufman, R.B. Griffiths, Phys. Rev. B {\bf 24}, 496 (1981).

\bibitem{Kaufman84} M. Kaufman, R.B. Griffiths, Phys. Rev. B {\bf 30}, 244 (1984).



\bibitem{Erbas} A. Erbas, A. Tuncer, B. Yucesoy, A. N. Berker, Phys Rev E {\bf 72},
026129 (2005).

\bibitem{Hinczewski} M. Hinczewski, A. N. Berker, Phys. Rev. E {\bf 73}, 066126 (2006).

\bibitem{Rozenfeld} H. D. Rozenfeld, D. ben-Avraham, Phys Rev E {\bf 75}, 061102(2007).

\bibitem{Kaufman84b}M. Kaufman, D. Andelman, Phys. Rev. B {\bf 29}, 4010-4016 (1984).

\bibitem{Hu} C.-K. Hu and C. N. Chen, Phys. Rev. B {\bf 38}, 2765 (1988).

\bibitem{Fortuin} C. M. Fortuin, P. W. Kasteleyn, J. Phys. Soc. Jpn. Supplm. {\bf 26}, 11
(1969).

\bibitem{Migdal} A. A. Migdal, JETP (SovPhys){\bf 42}, 743 (1976).

\bibitem{Kadanoff} L. P. Kadanoff, Ann. Phys.(NY) {\bf 100}, 359 (1976).

\bibitem{Kaufman83} M. Kaufman, R.B. Griffiths, Phys. Rev. B {\bf 28}, 3864 (1983).

\bibitem{Lindemann}  F. A. Lindemann, Z. Phys {\bf 11}, 609 (1910).

\bibitem{Kaufman84a} M. Kaufman, Phys. Rev. B {\bf 30}, 413(1984).

\bibitem{Chase}S. I. Chase, M. Kaufman, Phys. Rev. B {\bf 33}, 239-244 (1986).

\bibitem{Binder} K. Binder and D. W. Heermann, {\it
Monte Carlo Simulation in Statistical Physics}, Springer (2002).

\bibitem{Ferren} A. M. Ferrenberg and R. H. Swendsen, Phys. Rev. Lett.
{\bf{61}}, 2635(1988); Phys. Rev. B {\bf{44}}, 5081(1991).

\bibitem{Bergmann} D. J. Bergmann and B. I. Halperin, Phys. Rev. B {\bf 13}, 2145 (1976) and references therein.

\bibitem{Fisher68}  M. E. Fisher, Phys. Rev. {\bf 176}, 257 (1968).

\bibitem{Diep}  E. H. Boubcheur and H. T. Diep, J. Appl. Phys. {\bf 85}, 6085 (1999);
      E. H. Boubcheur , P. Massimino, H. T. Diep, J. of Magn. and Magn. Mater. {\bf 223}, 163-168 (2001) and references therein.

\bibitem{Landau}  Xiaoliang Zhu, D. P. Landau, and N. S. Branco,
Phys. Rev. B {\bf 73}, 064115 (2006).

\end{thebibliography}
\end{document}